*RaDMaX online*: a web-based program for the determination of strain and damage profiles in irradiated crystals using X-ray diffraction

A. Boulle [a] and V. Mergnac [b]

[a] Institut de Recherche sur les Céramiques, CNRS UMR 7315, Centre Européen de la Céramique, Limoges, France

[b] Université de Limoges, Direction des Systèmes d'Information, Limoges, France.

**Abstract**

*RaDMaX online* is a major update to the previously published *RaDMaX* (radiation damage in materials analysed with X-ray diffraction) software [Souilah, Boulle & Debelle, *J. Appl. Cryst.* (2016) **49**, 311-316]. This program features a user friendly interface that allows to retrieve strain and disorder depth-profiles in irradiated crystals from the simulation of X-ray diffraction data recorded in symmetrical θ/2θ mode. As compared to its predecessor, *RaDMaX online* has been entirely rewritten in order to be able to run within a simple web browser, therefore avoiding the necessity to install any programming environment on the users' computers. The *RaDMaX online* web-application is written in *Python* and developed within a *Jupyter* notebook implementing graphical widgets and interactive plots. *RaDMaX online* is free and open source (CeCILL license) and can be accessed on the internet at: https://aboulle.github.io/RaDMaX-online/.

**1. Introduction**

Together with Transmission Electron Microscopy, Raman spectroscopy and Rutherford back-scattering spectroscopy in channelling mode (RBS/C), X-ray diffraction (XRD) plays a crucial role in the characterization of irradiated materials (Zhang *et al.*, 2015). XRD distinguishes itself by the fact that it is sensitive to both lattice strain and atomic disorder, and in that it allows to derive these quantities on a an absolute scale. For instance, whereas Raman scattering and RBS/C quantify disorder through *ad hoc* parameters such as the broadening of the Raman lines or the fraction of displaced atoms determined from the the back-scattering yield, XRD in principle allows to determine the actual atomic displacement probability distribution (Boulle & Debelle, 2016), albeit this comes with an increased difficulty in the data analysis as compared to others techniques.

The analysis of XRD data does not necessarily have to be complex; for instance the determination of the maximum strain in the irradiated region can be easily obtained by measuring the position of the peak emanating from the strained region (Debelle & Declémy, 2010) and an estimation of the strained depth can be obtained by analysing the evolution of the width of the interference fringes (Sousbie *et al.*, 2006). However in order to benefit from the full potential offered by XRD, namely the retrieval of depth-resolved strain and disorder profiles, the simulation of the experimental data is mandatory (Zaumseil *et al.*, 1987; Klappe & Fewster, 1994; Milita & Servidori, 1995; Emoto *et al.*, 2009; Boulle & Debelle, 2010; Rieutord *et al.*, 2013).

In 2016, this led to the development of the computer program *RaDMaX* [Radiation Damage in Materials analysed with X-ray diffraction (Souilah *et al.*, 2016)] which, contrarily to most XRD simulation software packages, is solely dedicated to the simulation of XRD data recorded from irradiated crystals. This particular focus allows to significantly simplify the user interface, hence allowing scientists even with only little crystallographic knowledge to analyse their data. *RaDMaX* is written in Python which makes it compatible with all major computer operating systems (Windows, Mac OS and Linux). The downside of this multi-platform capability is that it requires the installation of a full scientific Python software distribution, which can be a disincentive for some users to adopt the software. This can happen, for example, when the users don't have administrator privileges on their computer, or when they are not familiar with setting up a computing environment. For this reason we engaged in developing a web-based version of *RaDMaX* that entirely runs within a web browser and, therefore, doesn't require the installation of any software, while keeping the possibility for scientists to load and analyse their own data and download complete simulation results.

This software, *RaDMaX online*, described in the present article (as well as the corresponding source code files), can be accessed at the following address: https://aboulle.github.io/RaDMaX-online/. As compared to the previous version of the program (Souilah *et al.*, 2016), the code has almost entirely been rewritten in order be able to be run on a web-hosted *Jupyter* server that the users can access *via* a simple html page. In section 2 we provide details regarding the implementation of the program (scientific background and the libraries and technologies used), whereas in section 3 we describe the user interface of the program, particular emphasis being laid on the new features that have been developed since the previous release the *RaDMaX* program.

## 2. Background and implementation

In this section we shall briefly give recall the scientific foundations on which *RaDMaX online* relies on; complete descriptions are provided in our previous articles (Boulle & Debelle, 2010; Souilah *et al.*, 2016). We will also outline the computing libraries that have been used, although we are not going to provide the details of the algorithms used since the source code is freely available online as free software (CeCILL licence).

## 2.1 Scientific background

In *RaDMaX online* the diffracted X-ray intensity is computed within the framework of the dynamical theory of diffraction from distorted crystals as first theorized by S. Takagi and D. Taupin (Taupin, 1964; Takagi, 1969) which is the state of the art approach for the analysis of irradiated crystals using XRD (Klappe & Fewster, 1994; Milita & Servidori, 1995). More specifically, we make use of the recursive solutions to the Takagi-Taupin equations provided by Bartels *et al.* (Bartels *et al.*, 1986). Within this approach, the irradiated region is divided into *N* sub-layers in which the composition, strain and disorder are assumed to be constant (typical values for *N* are within the 50-150 range). Consistently with earlier studies, the disorder is quantified through the Debye-Waller factor (Speriosu, 1981; Milita & Servidori, 1995; Zaumseil *et al.*, 1987) :

$$DW(z) = \langle exp\left[i\mathbf{Q}\delta\mathbf{u}(z)\right]\rangle \qquad (1)$$

where **Q** is the scattering vector ($Q = 4\pi \sin\theta/\lambda$ where θ is half the scattering angle and λ is the x-ray wavelength), δ**u**(*z*) corresponds to the random atomic displacement vector at the depth *z* below the surface, and the average <...> at depth *z* is performed in (*x,y*) planes parallel to the crystal surface. The Debye-Waller factor lowers the coherently diffracted intensity; for a perfect crystal, DW = 1, whereas for heavily disordered or amorphous materials, DW → 0. The strain has its usual meaning which, in the case of the θ/2θ scan recorded in the direction normal to the surface, correspond to the third diagonal element of the strain tensor $e_{zz}$:

$$e_{zz}(z) = \frac{d_z(z) - d_z^{th}}{d_z^{th}} \qquad (2)$$

where $d_z(z)$ is the inter-planar spacing of the diffracting crystallographic plane family measured at depth *z*, and the superscript *th* refers to the theoretical value.

Starting from the unirradiated region of the crystal (which corresponds to the scattering from a perfect single crystal), the total diffracted amplitude is constructed iteratively by combining the amplitudes diffracted and transmitted at each sub-layer interface, up to the surface. Determining the depth-resolved strain and disorder profiles therefore consists in determining the values of the strain and disorder within each sub-layer, which is hardly feasible without further constraints put on the system. In *RaDMaX online*, we make used of cubic B-spline functions to model the strain and disorder depth-profiles which allows to reduce the number of fitting parameters by a factor ~ 10 (Boulle & Debelle, 2010). Using cubic B-splines, both the strain and the disorder have the following form:

$$f(z) = \sum_{i=1}^{N_w^{S,D}} w_i^{S,D} B_{i,3}(z) \qquad (3)$$

where $N_w^{S,D}$ is the number of B-splines used to describe the strain ('S') and disorder ('D') profiles (typical values are in the 5-15 range), $w_i^{S,D}$ are the weights to be determined in the fitting procedure and $B_{i,3}(z)$ is the third-degree basis function (Boulle *et al.*, 2003). With this model, the number of fitting parameters is reduced from $2 \times N$ to $2 \times N_w$.

*2.2 Python implementation*

All crystallographic calculations are implemented using the Python programming language ([https://www.python.org/](https://www.python.org/)), together with the *NumPy* library ([https://numpy.org/](https://numpy.org/)) which introduces vectorized operations, *i.e.* operations implicitly working on all components of a vector (Oliphant, 2007; van der Walt *et al.*, 2011), and the *SciPy* library ([https://www.scipy.org/](https://www.scipy.org/)) which provides a wealth of scientific functions. Further details regarding crystallographic computing using the Python programming language can be found in (Boulle & Kieffer, 2019). The structure factor of the material is computed using the *xrayutilities* python package (Kriegner *et al.*, 2013) which, among other things, allows to read crystallographic information files (*cif*) which contain the description of the structure of the crystal (Hall *et al.*, 1991). It is here worth mentioning that *cif* files can be obtained from the crystallography open data base ([http://www.crystallography.net/cod/](http://www.crystallography.net/cod/)) which provides an open access to a large collection of crystal structures.

The user interface has been developed within a *Jupyter* notebook ([https://jupyter.org/](https://jupyter.org/)), a free and open-source interactive programming environment that runs within a web browser and that allows to gather the source code, the computational output (including plots, figures, *etc.*) as well as rich text within a single document (Perkel, 2018). Interactive widgets (*i.e.* sliders, editable text boxes, selection boxes, etc.) have been implemented using the *ipywidgets* library ([https://github.com/jupyter-widgets/ipywidgets)](https://github.com/jupyter-widgets/ipywidgets)). Central to *RaDMaX online* is the existence of interactive strain and disorder depth-profiles that can be graphically manipulated by the user so as to match the computed XRD curve with experimental data. This feature is brought to *Jupyter* using the *bqplot* library ([https://github.com/bloomberg/bqplot](https://github.com/bloomberg/bqplot)). In short, *bqplot* is a plotting library where every attribute of a plot can be programmatically modified and which allows for high level of user interactivity (data selection, data manipulation, *etc.*).

Finally, although *RaDMaX online* can be run locally on every user's computer by downloading the source code, its intended primary operating mode is to run as a web-hosted application. In such a case, for obvious security reasons, the users should not be able to execute arbitrary code on the web server. Since recently this is feasible thanks to the *voilà* library ([https://github.com/voila-dashboards/voila](https://github.com/voila-dashboards/voila)). When the URL of the *Jupyter* notebook is called, *voilà* launches the kernel for that notebook, runs all the cells and populates the notebook model with the outputs. After execution, all the source code cells are hidden and the notebook is converted to html, including JavaScript widgets which establish the connection to the kernel, and this page is served to the user. The corresponding result for *RaDMaX online* is given in Fig. 1.

Regarding web hosting, *RaDMaX online* is running on a Docker – enabled server hosted at university of Limoges that can be directly accessed at the following location: https://radmax.unilim.fr/. In brief, Docker (https://www.docker.com/) is a container technology that allows to easily distribute and deploy a program and all its dependencies by means of "images". Notice that this also allows to run the program locally if Docker is properly installed on the users computer. As the rest of *RaDMaX online,* the sources for the Docker deployment are free to use and can be downloaded from: https://github.com/aboulle/RaDMaX-docker.

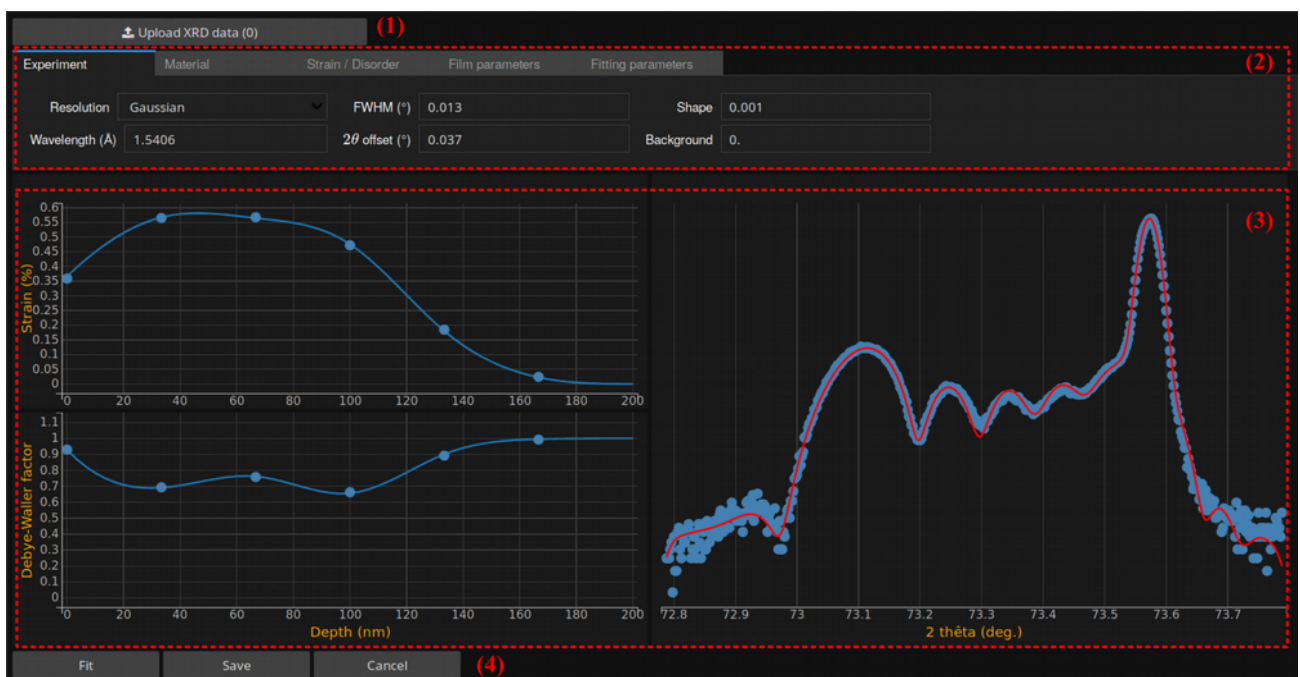

*Fig. 1. RaDMaX online's web interface.*

## 3. Program description and usage

As shown in Fig. 1, *RaDMaX online* is divided is 4 different areas:

1. An "upload" button allows the user to send their data to the server (in a two columns [$2\theta$ intensity] tab- or space-separated ascii format). It should be noted that the data is entirely stored in RAM and no data is saved on the server. By default *RaDMaX online* is loaded with a default data set corresponding to an irradiated (001)-oriented $ZrO_2$ single crystal

2. The different tabs ("Experiment", "Material", "Strain/disorder", "Film parameters" and "Fitting parameters") allow to set all parameters used in the XRD calculations and in the fitting procedure.

3. The experimental (filled circles) and calculated (continuous line) XRD curves are shown in the right-hand side plot, whereas the left-hand side plots are the strain and disorder depth-profiles. These plots are interactive and can be manipulated with the mouse via the control points (filled circles) so as to match the calculated curve with to experimental data.

4. Three buttons at the bottom allow to launch or cancel a least-squares refinement of the strain and disorder profiles (using the solution provided by the user as a starting point) as well as to download the simulation results.

*3.1 The "Experiment" tab*

As the name suggests, this tab allows to set the parameters related to the XRD experiment (Fig. 1). The first line ("Resolution", Full-width at half-maximum ("FWHM") and "Shape") correspond to parameters describing the angular resolution of the diffractometer. The exact angular resolution function is tightly linked to the type of optics that are used to condition the incident beam. A fast and efficient way to quantify the resolution is to use the the XRD peak from the (unirradiated) single crystal as a reference and to adjust the parameters of an arbitrarily chosen bell-shaped function so as to optimally describe the corresponding peak. In *RaDMaX online* several function are available: Gaussian, Lorentzian, pseudo-Voigt (*i.e.* a linear combination of a Gaussian and a Lorentzian function) and split (*i.e.* asymmetric) pseudo-Voigt. The width of the function is given by the FWHM parameter, whereas the "shape" parameter correspond to the relative amount of Gaussian (shape = 0) and Lorentzian (shape = 1) in the pseudo-Voigt function. Fig. 2 shows the calculated diffracted intensity around the 004 reflection of an irradiated (001)-oriented $ZrO_2$ single crystal. The corresponding strain depth-profile is given in Fig. 2a (the Debye-Waller is constant and equal to 1). Increasing the FWHM obviously increases the width of the Bragg peak emanating from the virgin region and lowers the fringes contrast (Fig. 2b), whereas increasing the shape parameter enhances the tails of the Bragg peak emanating from the virgin region and also decreases the fringes contrast (Fig. 2d).

This set of parameters might also be used to describe the scattering from imperfect single crystals. This is the reason why the possibility of asymmetric functions has also been implemented (a recent example can be found in (Boulle *et al.*, 2019)). This is achieved by selecting the split pseudo-Voigt function and by providing two, comma-separated values, for either or both the FWHM and shape values. These values correspond to left- and right-hand side parameters of the asymmetric function. The effect of a low-angle side asymmetry is shown in Fig. 2c. It can be observed that, in comparison with Fig. 2d, the difference in the diffraction emanating from the irradiated part is not obvious. Conversely, the asymmetry is clearly observed for the Bragg peak for the virgin area.

The second line allow to fix the x-ray wavelength, the background and a possible 2θ offset. Whereas the two former are rather self-explanatory, the latter correspond to an overall shift between the computed and the experimental curve. This offset can be due to a misalignment of the diffractometer or to the fact the material analysed differs from the *cif* file used for the calculation, resulting in a shift between the Bragg peak of the computed curve and the observed peak. The $2\theta$ shift resulting from a difference in lattice parameters is proportional to tan$\theta$; however, for sufficiently small angular ranges (~1° or less) the difference in lattice parameters can be safely corrected by applying a uniform 2$\theta$ shift.

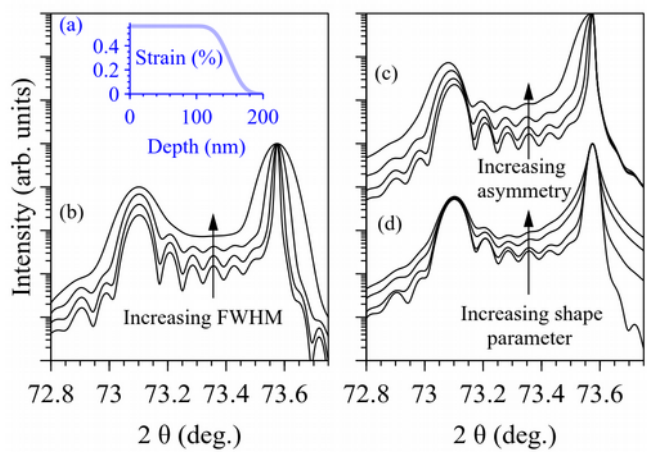

*Fig. 2.* (a) strain depth-profile used for all calculation in the present article. Evolution of the 004 reflection of a (001)-oriented ZrO$_2$ single crystal with (b) increasing FWHM of the resolution function (FHWM = 0.005°, 0.01°, 0.02° and 0.04°), (c) increasing asymmetry (FWHM left = 0.005°, 0.01°, 0.02° and 0.04°, FWHM right = 0.005°), (d) increasing shape parameter (shape = 0 [Gaussian], 0.1, 0.5 and 1 [Lorentzian]).

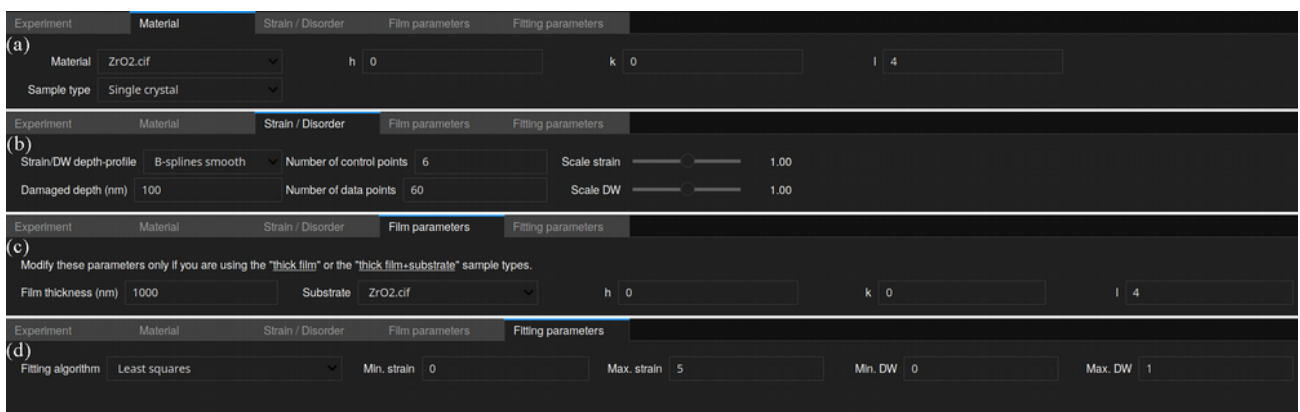

*Fig. 3.* The different tabs of *RaDMaX online's* interface.

## *3.2 The "Material" and "Film parameters" tabs*

The "Material tab", Fig. 3a, allows to select the material from a drop-down menu and to set the *hkl* values of the reflection. Selecting a material from the list computes the structure factor and lattice spacing for the corresponding *hkl* values, using *cif* files that are located within the 'structure' folder of the sources of *RaDMaX online*. Specific *cif* files can be added on demand or, when using *RaDMaX* in offline mode, added manually in the corresponding folder.

The "Sample type" drop-down menu allows to select the geometry of the irradiated crystal. Possible choices are "Single crystal", "Thick film", "Thick film + substrate" and "Thin film". These geometries are explicated below.

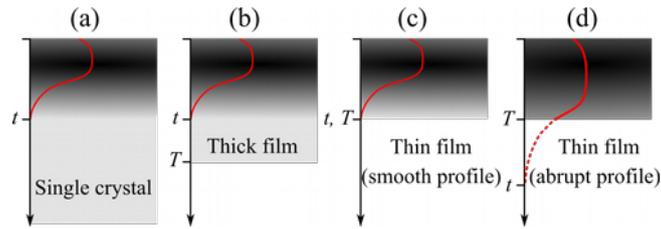

*Fig. 4.* The different sample geometries supported by *RaDMaX online*. $T$ is the thickness of the film or the substrate and $t$ is the depth of the damaged region. The shaded region schematizes the damaged region of the crystal. The red curve illustrates the depth-dependence of the damage (strain or disorder). The damage depth profile is determined by the energy loss of the incident projectiles within the crystal and therefore smoothly decrease to 0 (after one or more maxima) (a-c), except when the film thickness is smaller than the total projectile path length, in which case the damage is abruptly truncated (d).

Fig 4a, represents the case of a single crystal and the red curve correspond to the strain or disorder profile. The strain and disorder profiles are determined by the energy loss of the incident projectiles within the target material and, in general, exhibit a smooth transition between the irradiated and the non-irradiated regions. This smooth transition is enabled by selecting the "Smooth" profile in the "Strain/disorder" tab (see section 3.3). This option is selected by default. The irradiated thickness is referred to as $t$, whereas the thickness of the single crystal is referred to as $T$. For a sufficiently thick single crystal $T \to \infty$. The corresponding XRD curve using the same strain depth-profile as in Fig. 2a is shown in Fig. 5a (black curve).

For a thick film, $T$ is not infinite but larger than the irradiated thickness ($T > t$). The corresponding $T$ value can be given under the "Film parameters tab", Fig. 3c. Fig 5a (red curve) shows the XRD curve in the case of 500 nm thick $ZrO_2$ film with 200 nm irradiated region. If a substrate is diffracting in the recorded angular region its nature and the corresponding *hkl* can be selected under the same tab (Fig. 3c).

Fig. 4c is a special case of a thin film where the irradiated thickness exactly correspond to the film thickness, $t = T$. This situation is computed when the "Thin film" geometry is selected while keeping the "Smooth" strain profile (the same situation can be computed by selecting "Thick film" and setting $t = T$). The corresponding XRD data is given by the blue curve in Fig. 5a: the dome located at the $2\theta$ position corresponding to the perfect crystal is due to the narrow, low strain, region in Fig. 2a. Although not impossible, this situation seems rather far-fetched; a more realistic example where the film thickness is actually thinner than the expected projectile path within the material ($T < t$) is represented in Fig. 4d. This situation can be computed using the "Thin film" option combined with the "Abrupt" strain profile under the "Strain/disorder" tab. The calculated XRD curve is shown as the green line in Fig. 5a, where a perfect thin film signal is easily recognized.

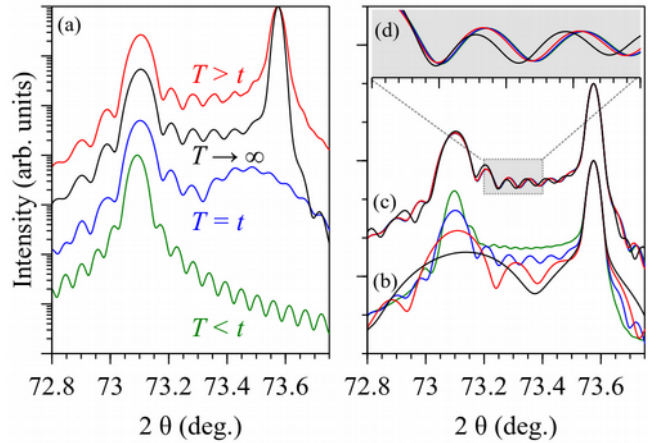

*Fig. 5.* (a) calculated XRD curves corresponding to geometries depicted in Fig. 4. Single crystal (black), thick film (red), thin film smooth (blue) and thin film abrupt (green). (b) calculated curves for increasing damaged depths: 50 nm (black), 100 nm (red), 200 nm (blue) and 400 nm (green). (c) calculated XRD curves for an increasing numbers of sub-layers used in the computation: 10 (black), 30 (red), 60 (blue) and 100 (green) sub-layers. (d) zoom on the 73.2 – 73.4° region.

## *3.3 The "Strain/disorder" tab*

This tab allows the set all parameters related to the strain and disorder profiles (Fig. 3b). The "Strain/DW depth-profile" allows to select either the "Smooth" or the "Abrupt" B-spline functions, as discussed in section 3.2 above. The "Number of control points" correspond to the number of interactive points in the strain and in disorder profiles; from a numerical point of view, this parameter correspond to the $N_w$ constant used in Eq. 3. This value should be adjusted to the complexity of the strain/disorder profile as already discussed in (Boulle & Debelle, 2010; Souilah *et al.,* 2016) . Low values (say, < 10) are well adapted to rather simple profiles, whereas complex profiles (multimodal or exhibit tiny features) may require larger values (> 10). It should be borne in mind that the number of fitting parameter is equal to $2N_w$; hence, increasing $N_w$ value might yield to more unstable least-squares fitting procedures.

The "Damaged depth", as discussed in section 3.3, is the depth required for the strain or the disorder to converge to their bulk values (0 and 1, respectively). Fig. 5b shows the evolution of the XRD curve when the damaged depth takes the following values: 50 (black), 100 (red), 200 (blue) and 400 nm (green). Obviously, decreasing the damaged depth leads to broader profiles and more widely spaced interference fringes. This feature can be used to estimated the damage depth from the

XRD signal.

The "Number of data points" text field correspond to the number of sub-layers used in the calculation of the XRD intensity (the parameter $N$ mentioned in section 2.1). This parameter is not straightforward to determine. Fig. 5c and 5d show the influence of this parameters when it takes the following values: 10 (black), 30 (red), 60 (blue) and 100 (green). It can be noticed that no visible difference can be observed between the 60 and 100 sub-layers cases. A rule of thumb is that 10 to 20 sub-layers should be used for the tiniest feature in the strain/disorder profile. In the case of the strain depth-profile show in Fig. 2a, the abrupt decrease of strain ranges from 120 to 180 nm, that is a 60 nm thickness, which, with above criterion gives a sub-layer thickness of 3 to 6 nm, which in turn gives a total number of sub-layers ranging between 33 to 66 in perfect agreement with the observation of Fig. 5c,d. Increasing the value of $N$ increases the depth resolution but increases the computation time; conversely decreasing this value accelerates the computation at the expense of the depth resolution.

Finally, two sliders can be used to apply a global scale factor to the strain and the Debye-Waller values, where the latter is constrained to lie within the [0,1] interval.

*3.4 The "Fitting parameters" tab*

This last section of the user interface allows choose the least-squares fitting algorithm and the permitted variation range of the fitted parameters. The default algorithm is the "trust reflective algorithm" as implemented in *SciPy* (Branch *et al.*, 1999). Regarding the bounds set on the parameters, it is recommended to keep the Debye-Waller factor within its range of physically meaningful values, *i.e.* the [0,1] interval. The bounds on the strain are dependent on the problem and should be determined by the user. It should be noted that the limits can be ignored by selecting the "Least squares (no bounds)" fitting algorithm. This implements the more usual Levenberg-Marquardt algorithm.

Finally, contrarily to previous versions of *RaDMaX*, this version does not include the more computationally demanding simulated annealing fitting procedure in order not to saturate the resources of the web server.

**4. Alternate working modes**

As mentioned earlier *RaDMaX online* can be accessed on the web. The main address is: https://radmax.unilim.fr/. If, for any reason, the main server is not available, an alternate solution consist in running the *RaDMaX* notebook on the *Binder* web service (https://mybinder.org/). In short, this service allows to connect to a software repository (like github, gitlab, *etc.*) and to run the content within a *Jupyter* server that is build on demand, including all the necessary dependencies.

The drawback is that building the image of the custom *Jupyter* server might take some times (up to a few minutes) and that, being a free service, the images can not be kept indefinitely on Binder and are therefore deleted after a given period of inactivity. Nonetheless, this service remains a very convenient way to share scientific notebooks and even complete web applications like *RaDMaX online*. The corresponding binder link is the following:

https://mybinder.org/v2/gh/aboulle/RaDMaX-online/master?urlpath=voila/render/RaDMaX.ipynb

Advanced users might want to customize *RaDMaX* to fit their own needs. This is feasible by downloading the source code from https://aboulle.github.io/RaDMaX-online/ and running the RaDMaX.ipynb file in their own *Jupyter* installation. For users having installed their Python environment though the Anaconda distribution (https://www.anaconda.com/distribution/) *Jupyter*, as well as *SciPy* and *NumPy* are already included. The only remaining dependencies to be installed are ipywidgets, bqplot and xrayutilities. The last one can only be installed with the Python package manager. The following should be typed in a command-line interface:
```
pip install xrayutilities
```
The others can be installed either via pip or via anaconda's package manager, *i.e.*
```
pip install ipywidgets bqplot voila
```
or
```
conda install -c conda-forge ipywidgets bqplot voila
```
If the Python environment has been installed without using Anaconda, the *NumPy*, *SciPy* and *Jupyter* packages should be installed as well. There are many ways to do that depending on the operating system; the following command should work in any environment:
```
pip install numpy scipy jupyter
```

## 5. Concluding remarks

*RaDMaX online* is a free software dedicated to the determination of strain and disorder depth-profiles in irradiated crystals by means of XRD. It is a web-based program that doesn't require any installation on the users' computers. *RaDMaX online* is entirely based on free and open-source technologies. The program as well as the source code can be accessed at the following address:
https://aboulle.github.io/RaDMaX-online/
The web application can be directly accessed at: https://radmax.unilim.fr/

**Acknowledgements**

AB is grateful to the NEEDS program (MeSINII project) for partial support of this work.

**Figure captions**

Fig. 1: *RaDMaX online*'s web interface.

Fig. 2: (a) strain depth-profile used for all calculation in the present article. Evolution of the 004 reflection of a (001)-oriented $ZrO_2$ single crystal with (b) increasing FWHM of the resolution function (FHWM = 0.005°, 0.01°, 0.02° and 0.04°), (c) increasing asymmetry (FWHM left = 0.005°, 0.01°, 0.02° and 0.04°, FWHM right = 0.005°), (d) increasing shape parameter (shape = 0 [Gaussian], 0.1, 0.5 and 1 [Lorentzian]).

Fig. 3: The different tabs of *RaDMaX online's* interface.

Fig. 4: The different sample geometries supported by *RaDMaX online*. *T* is the thickness of the film or the substrate and *t* is the depth of the damaged region. The shaded region schematizes the damaged region of the crystal. The red curve illustrates the depth-dependence of the damage (strain or disorder). The damage depth profile is determined by the energy loss of the incident projectiles within the crystal and therefore smoothly decrease to 0 (after one or more maxima) (a-c), except when the film thickness is smaller than the total projectile path length, in which case the damage is abruptly truncated (d).

Fig. 5: (a) calculated XRD curves corresponding to geometries depicted in Fig. 4. Single crystal (black), thick film (red), thin film smooth (blue) and thin film abrupt (green). (b) calculated curves for increasing damaged depths: 50 nm (black), 100 nm (red), 200 nm (blue) and 400 nm (green). (c) calculated XRD curves for an increasing numbers of sub-layers used in the computation: 10 (black), 30 (red), 60 (blue) and 100 (green) sub-layers. (d) zoom on the 73.2 – 73.4° region.

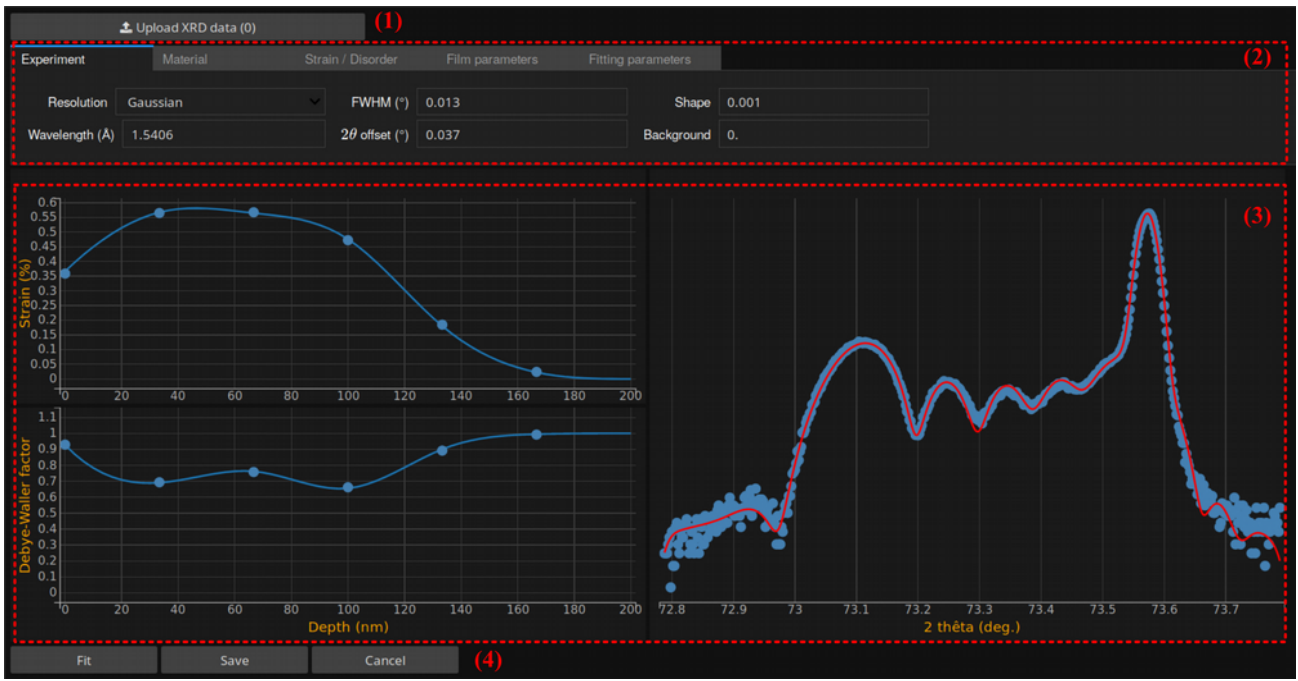

Fig. 1

*RaDMaX online*'s web interface.

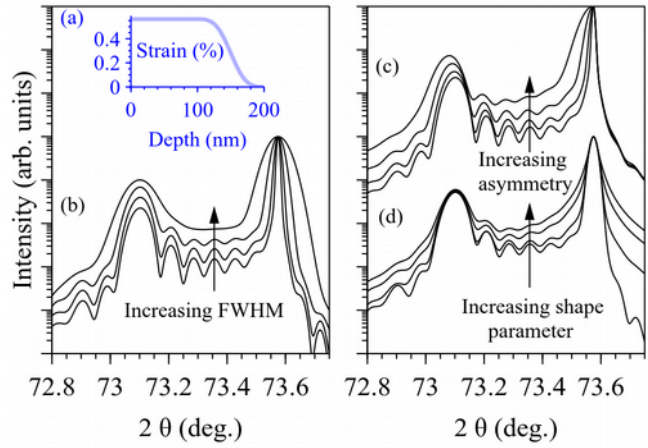

Fig. 2

(a) strain depth-profile used for all calculation in the present article. Evolution of the 004 reflection of a (001)-oriented $ZrO_2$ single crystal with (b) increasing FWHM of the resolution function (FHWM = 0.005°, 0.01°, 0.02° and 0.04°), (c) increasing asymmetry (FWHM left = 0.005°, 0.01°, 0.02° and 0.04°, FWHM right = 0.005°), (d) increasing shape parameter (shape = 0 [Gaussian], 0.1, 0.5 and 1 [Lorentzian]).

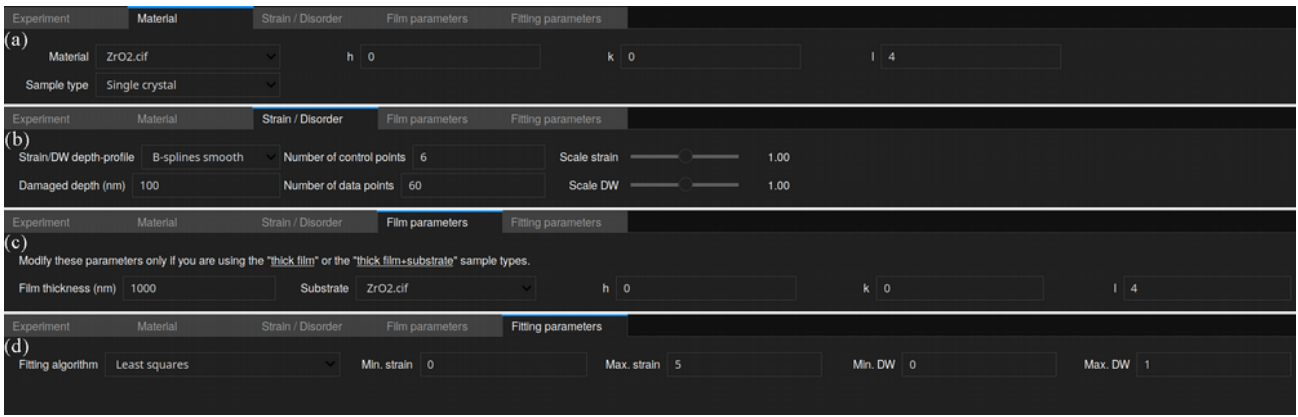

Fig. 3

The different tabs of *RaDMaX online's* interface.

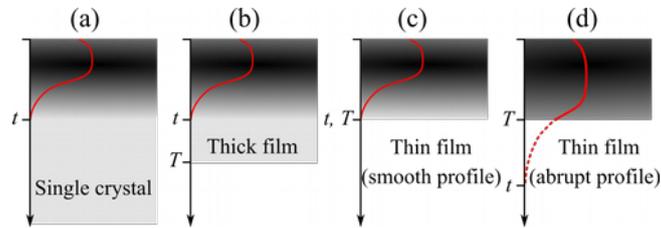

Fig. 4

The different sample geometries supported by *RaDMaX online*. *T* is the thickness of the film or the substrate and *t* is the depth of the damaged region. The shaded region schematizes the damaged region of the crystal. The red curve illustrates the depth-dependence of the damage (strain or disorder). The damage depth profile is determined by the energy loss of the incident projectiles within the crystal and therefore smoothly decrease to 0 (after one or more maxima) (a-c), except when the film thickness is smaller than the total projectile path length, in which case the damage is abruptly truncated (d).

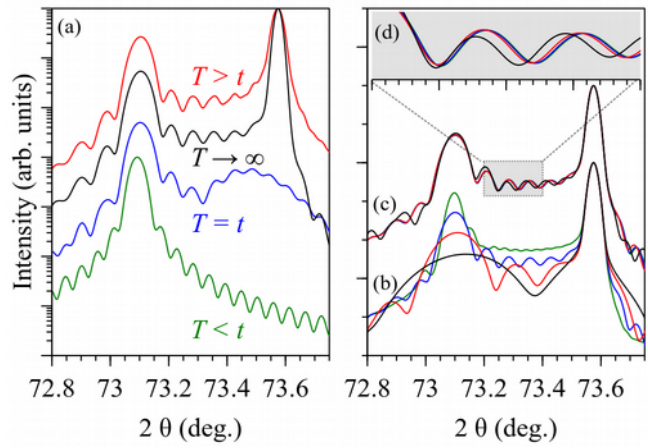

Fig. 5

(a) calculated XRD curves corresponding to geometries depicted in Fig. 4. Single crystal (black), thick film (red), thin film smooth (blue) and thin film abrupt (green). (b) calculated curves for increasing damaged depths: 50 nm (black), 100 nm (red), 200 nm (blue) and 400 nm (green). (c) calculated XRD curves for an increasing numbers of sub-layers used in the computation: 10 (black), 30 (red), 60 (blue) and 100 (green) sub-layers. (d) zoom on the 73.2 – 73.4° region.